\documentclass[preprint,1p,authoryear]{elsarticle}

\usepackage{amssymb}

\usepackage[utf8]{inputenc}

\usepackage{dirtytalk}
\usepackage{graphicx}
\usepackage{subfig}
\usepackage{times}
\usepackage{epsfig}
\usepackage{capt-of}
\usepackage{latexsym}
\usepackage{amsmath}
\usepackage{setspace}
\usepackage{epsfig}
\usepackage{multirow}
\usepackage{url}
\usepackage{fixltx2e}
\usepackage{mathtools}
\usepackage{color}
\usepackage{todonotes}
\usepackage{makecell}

\DeclareUnicodeCharacter{202C}{}

\journal{Information Processing \& Management}

\begin{document}

\newcommand{\longedge}[3]{{#1}\buildrel{#2}\over\longrightarrow{#3}}
\begin{frontmatter}



\author[addr1]{Saeed Amal\corref{cor1}\thanks{tnx}}
\ead{samal@is.haifa.ac.il}
\author[addr1]{Einat Minkov}
\ead{einatm@is.haifa.ac.il}
\author[addr1]{Tsvi Kuflik}
\ead{tsvikak@is.haifa.ac.il}
\address[addr1]{{\it Department of Information Systems, University of Haifa}}
\cortext[cor1]{Corresponding author}

\title{Person Entity Profiling Framework: Identifying, Integrating and Visualizing Online Freely Available Entity-Related Information}


\author{} 

\address{}

\begin{abstract}
When we consider our CV, it is full of entities that we are or were associated with and that define us in some way(s). Such entities include where we studied, where we worked, who we collaborated with on a project or on a paper etc. Entities we are linked to are part of who we are and may reveal about what we are interested in. Hence, we can view any CV as a graph of interlinked entities, where nodes are entities and edges are relations between them. This study proposes a novel entity search framework that in response to a real-time query about an entity, searches, crawls, analyzes and consolidates relevant information that is freely available on the Web about the entity of interest, culminating in the generation a profile of the searched entity. Unlike typical entity search settings, in which a ranked list of entities related to the target entity over a pre-specified relation is processed, we present and visualize rich information about the entity of interest as a typed entity-relation graph without an apriori definition of the types of related entities and relations. This view is structured and compact, making it easy to understand as well as interpret. It enables the user to learn not only about the entity in question, but also about related entities, thereby obtaining a better understanding of the entity in question. We evaluated each of the framework’s  components separately and then performed an overall evaluation of the framework, its visualization and the interest of users in the results. The results show that the proposed framework performs entity searches, related entity identification and relation identification very well and that it satisfies users' needs.

\end{abstract}


\end{frontmatter}


\section{Introduction}
\label{sec:intro}

Persons, organizations, locations and other entities to which we may be linked reflect our biographies and interests. The organizations in which we have studied and worked for, the people with whom we have collaborated, papers we have authored, and events in which we have participated are all examples of entities that represent our personal biography and interests. Thus, when presented with an unknown person entity or wanting to know something about such an unknown person, a user who accesses and peruses such information will get a good idea who this person is. 

\begin{figure*}[t]
\centering
\centering
\begin{small}
\includegraphics[width=10cm]{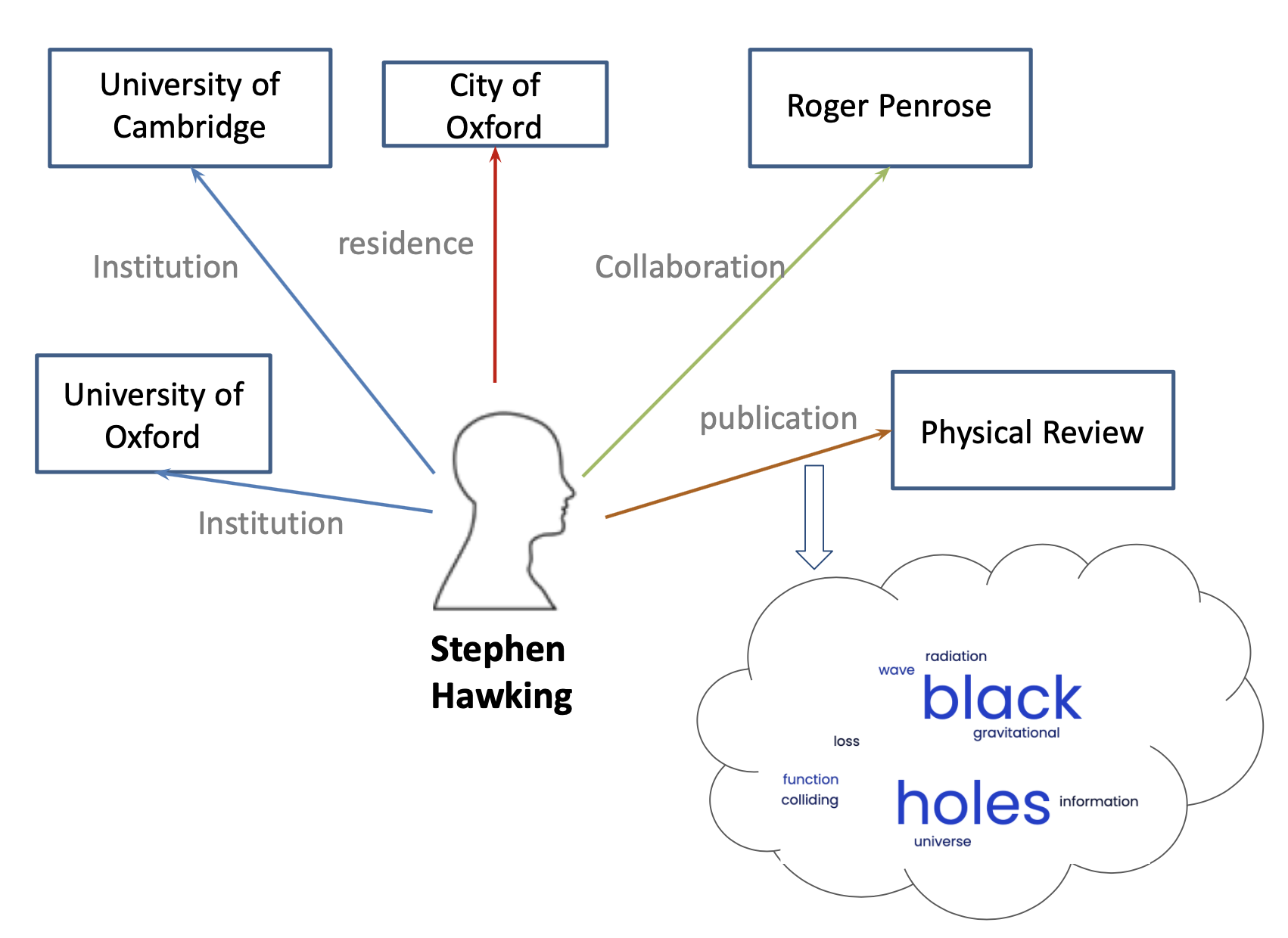} 
\caption{Example of an entity-based biography }
\label{fig:entitybasedbib}
 \end{small}
\end{figure*}

As an illustrative example, let us assume that a user is interested in retrieving information about a personality such as Stephen Hawking. Admittedly, relational facts about Hawking, including related entities and the nature of his relationships with them, is available in knowledge bases such as Wikidata\footnote{www.wikidata.org}. Some key relational facts are also presented by search engines such as Google\footnote{www.google.com} in the form of structured info boxes ~\citep{botaCHIIR16}; ~\citep{wuWWW08}. As illustrated in Figure~\ref{fig:entitybasedbib}, such relational facts corresponds to a typed graph. As shown in the graph,\footnote{This graph also includes a sample of the facts available in Wikidata about Hawking.} the person Stephen Hawking is linked to entities and concepts such as the ‘University of Oxford’, the ‘University of Cambridge’, ‘city of Oxford’, ‘Roger Penrose’ and ‘Physical Review’ (a scientific journal). The graph edges are directed and associated with relation types and word clouds associated with them, explaining the relation. The example graph was drawn based on the Wikipedia\footnote{www.wikipedia.org} page for "Stephen Hawking". Resources such as Wikipedia pages, however, are not available for most people. This paper offers a method for constructing entity profiles from raw data extracted from any relevant Web pages (e.g., personal homepages and other highly relevant Web pages, such as articles and blogs about the target entity). Having constructed such a personal profile, the answers to other complex questions such as: "What is the social network of person X?", "What is the relation type between persons X and Y?", and more, can be sought. Such information may be of interest, for instance, when someone applies for a position or public office when additional information beyond what is available in their CV or homepage is needed.
The entity-focused search approach we set out to develop includes the following features: 
\begin{itemize}
    \item Automatic construction of an entity profile that consolidates information from multiple relevant URLs, thus providing a comprehensive overview about the entity in question by describing their relations with other entities in the world.
    \item Support of high-level operations, including finding relations in the graph between two given entities that are not directly connected, or producing a ranked list of associated entities.
\end{itemize}

In this paper, we propose a framework for addressing the entity search challenge using information that is freely available on the Web, analyzing it and visualizing the results as a linked entity graph (the entity profile). There were three main tasks that achieving this goal presented to us were:
\begin{itemize}
    \item Identification of relevant Web pages
    \item Extraction of named entities and their typed relations with the target entity, from semi-structured and unstructured text 
    \item Visualization of the resulting entity profile.
\end{itemize}

The paper presents in details a complete work and its evaluation while initial ideas and results were already presented in different venues and forms during the development of the framework. Here we present the contribution beyond what was already published. 

The idea for harvesting and presenting entity-based profiles was initially presented in ~\cite{amal2017harvesting}, where positive feedback was provided by academics.Then a framework based on the idea and a demonstration of its application to the academic domain was presented in ~\cite{amal2019relational}. This paper used a preliminary implementation of the three above-mentioned tasks. The present paper offers a far more advanced version of the initial methodology. in the earlier stages, for the first task, we used a technique that identifies homepages only whereas here we use a more general technique that automatically identifies any relevant Web pages and we performed a more thorough evaluation on a larger scale dataset. Regarding the second task, for the extraction of named entities subtask, although the techniques then and now are the same, here we performed a massive evaluation on larger scale datasets and on a variety of domains, beyond the academic ones previously used. For the typed relation extraction subtask, here we proposed new techniques that involve deep learning and evaluated them on larger a scale dataset and on larger variety of relation types. For the third task, in the above-mentioned paper for the academic domain, we used a simplistic text-based visualization while focusing on users' perception of the idea. 

In addition, as the visualization of the results appeared to be highly important, being what the end user sees, this aspect was the focus of several studies. Initial results were presented in ~\cite{amal2019enhancing} and improved incrementally over several iterations and demonstrated in ~\cite{amal2020demonstrating}; ~\cite{umapamal2020}; ~\cite{amal2020personalized}, all of which used the academic domain for demonstration. 

To summarise, the present paper provides an integrated, in-depth presentation of the framework and its evaluation, above and beyond the initial results presented in earlier studies. We provide a detailed description of the different components, a detailed description of the experimentation and the evaluation of the individual components, also in a wider scope by considering several domains. Additionally, we present a novel visualization, developed as a result of feedback received in prior studies.   
The results show that the proposed framework (and the system implementing it) performs well in entity search (an F1 of 0.92 in identifying relevant pages, finding more than 20\% related entities, beyond what can be found in people's homepages and an F1 of ~0.9 in relation identification) and that it satisfies users’ needs effectively(the visual features were rated 3.7 on average on a scale of 1 to 5).

The rest of the paper is organized as follows: Section 2 presents related work in the areas of personal profile extraction from the Web, entity search and graph visualization. Section 3 presents our graph-based entity profiling and visualization framework and its components. Section 4 offers a set of experiments designed to evaluate the individual components and the whole framework. Section 5 discusses the implications and limitation of the study and its results. Finally, Section 6 concludes and suggests directions for future research.

\section{Related Work}
Our work integrates several aspects - from entity search and profile extraction to graph visualization. This section reviews related work and explains what we adopted and followed and in what ways our work differs from the related work.
\label{sec:related}
\subsection{Personal Profile Extraction from the Web}
Our work is highly related to ~\cite{chamoso2020} who presented a system for retrieving personal information from the Internet matching several input criteria while focusing on information in Spanish about people living in Spain. Their system searches for a person’s personal information mined by search engines such as Google or Bing and uses this information to create a personal profile. The input criteria for performing the search include a person’s full name and a series of keywords associated with them. The information is retrieved from multiple sources and analysed using a set of well‐known AI methodologies such as TF-IDF, SVMs,CNN. According to the authors, theirs is the only system that consolidates  all this data in one place. Once the system gathers the information, it is visualized as an entity card in a template with known tags. While Chamoso et al. (2020) focus on gathering information from sources containing information in Spanish about people living in Spain, our work is not specific to a language or persons from specific countries. In addition, our input is the person's name while they provide a series of keywords. 

Our work is also related to a recent work, WISER, presented in ~\cite{wisertool}. WISER generates entity-based profiles of scholars in Wikipedia and uses these profiles to explore the work and careers of academic experts. It indexes each academic author by applying their proposed  profiling technique that models the respective academic’s expertise with a small, labeled and weighted graph drawn from Wikipedia. Nodes in this graph are the Wikipedia entities mentioned in the academic’s publications, whereas the weighted edges express the semantic relatedness among these entities. The relatedness is computed by textual and graph-based relatedness functions, which are applied on the a small weighted subgraph in Wikipedia that consists of the Wikipedia graphs of the two entities. Then for each author, every node in the graph is labeled with a relevance score that models the pertinence of the corresponding entity to the academic’s  expertise, and is computed by means of a proper random-walk calculation over their graph. At query time, WISER seeks to identify the areas of expertise mentioned in the input query q and then retrieve a set of candidate experts to which it assigns an expertise score. This score is used for generating the final ranking of experts that are returned as result of query q. In our case, our system is more generic. Entities are not limited to scholars but can be any entity of type {\it person} and can be found in any unstructured Web page and not restricted to entities appearing in structured pages such as Wikipedia. 

Our work is also highly related to the work of ~\cite {Gysel2016}, where the main emphasis is on unsupervised profile construction, improving the efficiency of the query by performing query expansion with nearby terms based on traditional language models, and on semantic matching between query terms and candidate profiles. Their system learns a profile-centric latent representation of academic experts from their publications. While the learned profile is publication-based, we collect any information that seems relevant to the user and hence our approach can be generalized to domains other than academic researchers, as we demonstrate later. In addition, the work of ~\cite{Gysel2016}  is limited to the use of latent concepts, namely, ones that cannot be explicitly described and thus the system cannot explain why an expert profile matches a user’s query, while our work has several visual layers of information, including the sentence and word cloud of the main concepts connecting the user to the other related entities and the relation type as well as temporal aspects, making the result easily explainable. According to ~\cite{chamoso2020}, while there are well‐known AI methodologies that have been widely used for each component of profile construction, there is no published research that groups all of them together to build a profile.\\

\subsection{Entity Search}

Our work is related to entity search, known also as entity retrieval ~\citep{fissaha} or entity extraction ~\citep{pennacchiotti2011machine}. In entity search, the output of the search engine is a list of Web pages, ranked according to their relevance to the target entity ~\citep{pennacchiotti2011machine}. Several entity-focused search tasks have been defined in recent years by the IR community, including: 1. Entity Ranking (ER) ~\citep{inex2007}, where in this task, the query describes a topic of interest in natural language and specifies a target relation of related entities. In response to the query, a ranked list of entities is returned, as presented in their Wikipedia pages (here it is assumed that all entities have a Wikipedia page). 2. Related Entity Finding (REF) ~\citep{balog2010}, where in this task, defined by the Text REtrieval Conference (TREC) in 2009, an entity of interest is specified by its name or its homepage, and the goal is to retrieve entities that are related to the target entity over a specific relation type that is descried in natural language. 3. Entity List Completion (ELC) ~\citep{balog2010}, which, similarly to REF, focuses on finding entities that are associated over a specific relation with a target entity. In this task, unlike the REF task where relation is described in natural language, the desired relation should be concluded according to the relations to a given small set of example relevant entities. All these tasks are limited to pre-defined and specific types of relations, and assume that the query and target entities appear in Wikipedia, or in collections such as DBpedia or ClueWeb, or that their homepages are given. While the three above-mentioned entity search approaches return a ranked list of entities to represent the entity in query, we represent an entity of interest as a network of related entities and relationships without limiting the search to structured sources.

Our work is related to ~\cite{Nikolaev2020} who presented the Knowledge graph Entity and Word Embedding for Retrieval (KEWER) method, which embeds entities and words into the same low-dimensional vector space and takes into account a knowledge graph’s local structure and structural components, such as entities, attributes, and categories. It is designed specifically for entity search. KEWER is based on random walks over the knowledge graph and can be considered a hybrid of word and network embedding methods. It utilizes contextual co-occurrences as training data while treating words and entities as different objects. KEWER takes into account a knowledge graph's local structure. According to the authors, KEWER outperformed DBpedia-Entity v2 (DEv2), which is the standard test collection for entity search ~\citep{Hasibi17}.

Our work is also related to ~\cite{esmeir-2021-serag} who proposed Semantic Entity Retrieval from Arabic knowledge Graphs (SERAG), which was built on the KEWER system. For each entity, the system uses DBpedia to generate both a document formed of directly linked textual information and a set of random walks starting at that entity. Given a query, BM25 ~\citep{zhiltov15} scores entities based on the relevance of their documents to the query. Then to enhance the ranking, a word2vec ~\citep{mikolov2013} model is built using random walks where each walk consists of entities and predicates along a path in the Arabic DBpedia graph. Each walk is considered a sentence and all sentences are concatenated into one corpus, used to train the word2vec model based upon which the related entities are ranked. To evaluate SERAG, the author created and shared a Modern Standard Arabic (MSA) version of DEv2.

Our work is also related to ~\cite{Torres2019} who introduced a two-stage method to conduct entity search over DBPEDIA. The first stage queries DBPEDIA using a short text phrase that aims to find, for one particular entity of interest, a list of entities where the former has the highest relevance and the latter are entities related in some degree to the first one. They do this using a conventional informational retrieval (IR) model, which employs classical similarity methods between queries and documents representing the entities, where the title field of the document is used as a representation of the entity of interest. The second stage is a graph diffusion process using a heat diffusion kernel known as heatrank ~\citep{Yang2007}, which is a variation of PageRank. The authors showed that graph diffusion can increment the performance of the retrieval ranking process of related entities.

Additionally, our work is related to ~\cite{Lin2018} who investigated the task of ad hoc entity retrieval from a knowledge graph with hierarchical entity types and entity descriptions aiming at improving the entity retrieval task. They developed a model that combines entity types and descriptions with the Fielded Sequential Dependence model (FSDM) ~\citep{zhiltov15}. Their model incorporates the types of the entities as well as their hierarchical sequence as additional context. In this way, they traverse the hierarchy and retrieve the entities with the maximum score of paths as the most relevant context for the searched entity. 
While the previously mentioned works base their entity search on searching knowledge graphs, we crawl, identify and consolidate freely available information from the Web including unstructured data.

Our approach resembles exploratory search ~\citep{yogev12}, having related entity suggestions presented in response to a query. It has been shown that entity cards (~\citep{botaCHIIR16},~\citep{yogev12}), structured information objects presented by search engines that highlight relationships between different entities associated with a given query, support exploratory search and increase user’s engagement ~\citep{botaCHIIR16}. The information displayed in entity cards is typically deduced and selected from available structured knowledge bases ~\citep{kangWWW11}, ~\citep{miliWWW15}. In contrast, we apply information extraction techniques to identify a variety of entities and relation types mentioned on the Web in the context of the entity of interest. Such automatic processing is necessary to retrieve unstructured information about persons (or entities of other types) for whom a Wikipedia page or a knowledge base entry are not available.

Our work is closely related to the work of ~\cite{adamic03} who mined student homepages to create a social network. They too extracted named entities from students’ homepages, and connected individuals based on common named entities as well as based on additional information sources such as mailing lists. Our work differs in terms of the goals and the methods we apply. ~\cite{adamic03} predicted whether a person was a friend of another based on the number of linked items that they had in common. According to them, student homepages provide a glimpse into the social structure of university communities and these relationships in context. Our focus is on entity profiling as opposed to community detection. We construct a semantically richer graph for this purpose, consisting of directed and type-labeled edges.

Our work is also related to searching relevant unstructured Web sources. ~\cite{McKeown2016} presented InfoScout, an interactive person search tool, which handles the task of retrieving relevant Web pages by presenting a list of document screenshots in response to a query, and requesting relevance judgements from the user. The user feedback is then used to re-rank the results. This work requires feedback from the user on every single query or assumes that examples of relevant Web pages or part of them are given, while we do not assume having any Web page as a starting point.
~\cite{Hasibi17} introduced Nordlys, a toolkit for entity-oriented cataloging, entity retrieval, entity linking, and target type identification of entities in DBpedia. Their work is limited to entities that exist in DBpedia, whereas our approach also is applicable to entities outside of Wikipedia. ~\cite{zhiltov15} proposed a fielded sequential dependence model for adhoc entity retrieval, and also relies on structured documents as information source. Nevertheless, there exist many unstructured Web pages with relevant information; finding them and extracting this information requires a different approach. The authors tested their system on DBpedia, hence it is also hard to compare our results to theirs.

To the best of our knowledge, there are no recent studies that identify highly relevant unstructured documents about a target entity, even though, according to the International Data Corporation, in 2018, 80\% or more of data found in the digital universe was unstructured. The company estimated that unstructured data will be 95\% of the global data in 2020. According to ~\cite{adnan2019analytical}, Traditional information extraction (IE) systems cannot deal efficiently with this huge deluge of unstructured big data. In this work, we reach relevant information about the target entity beyond what can be found in structured data, by identifying highly relevant unstructured Web pages about the entity of interest using machine learning to train a classifier for this task. Our classifier only has the raw content of the Web pages and tags of named entity mentioned within these Web pages, if available. Hence, we identify relevant Web pages beyond the entity’s homepage, thus providing a user with a more comprehensive overview of the entity of interest.\\

\subsection{Graph Visualization}
In addition to various aspects of entity search, our work is also related to the use of graph visualization techniques for information visualization. In recent years, efficient algorithms for layout computation and graph visualization metaphors were developed and the impact of the resulting visualizations on readability and task performance were studied. Moreover, recently methods that scale to several million nodes and edges were developed and hence the focus has shifted to the visual complexity and  human interpretability of the resulting layouts ~\citep{eades2015}.

Our work is highly related to ~\cite{chamoso2020} as described above. Once their system generates the information, they visualize it as an entity card in a template with known tags. While their work visualizes the person's personal information using entity card shapes with known tags, we visualize it in a more comprehensive way that allows easier exploration of the related entities and the presented information. 
~\cite{dark12} introduced the PivotPaths interface that exposes faceted relations as visual paths in arrangements that invite the viewer to ‘take a stroll’ through an information space. PivotPaths supports pivot operations as lightweight interaction techniques that trigger gradual transitions between views. They designed the interface to allow for casual traversal of large collections in an aesthetically pleasing manner that encourages exploration and serendipitous discoveries. While this work also visualizes the related entities and relationships as nodes and edges in a graph, we go beyond and present explanations of their relations by showing the context, which improves the transparency and explainability of the graph.

~\cite{yogev17} suggested that to expose relationships while a maintaining relevance-based layout of the search results, presentation of search results could benefit from using a two-dimensional layout instead of the conventional one-dimensional ranked list. To this end, they introduced HiveRel, a search system that presents search results as tiled hexagons on a map-like surface with center-out relevance ordering, in which relationships are displayed on-demand. While our approach also uses two-dimensional visualization instead of the one-dimensional option for ranked lists of related entities as they did in this work, we offer more by employing word cloud visualization to provide an explainable and more detailed overview about the found relationships and not only the relation type. In addition, we present the full context from where the relation was extracted.

~\cite{Schmauder15} investigated the problem of visualizing time-varying graphs as node-link diagrams having a specific layout by exploiting the links as timelines. Partially drawn links are used to show the graph dynamics by splitting each link into as many segments as time steps that have to be represented. They used conventional 2D layout algorithms while showing the evolution over time. They also used color-coded links to represent the changing weights and tapered links to reduce possible overlaps at the link target nodes, which would occur when using traditional arrow-based directed links. We are also interested in the challenge of the visual complexity involved in the temporal aspect; however, our primary focus is on the challenge of embedding more facets in our graph to provide more comprehensive information in the searched entity profile. To emphasize on the presentation and visualization of the generated user profiles, we use a 2D graph while providing comprehensive explanations using word cloud visualization on the edges of the graph.

\section{Graph-Based Entity Profiling and Visualization Framework}
In this section we provide a detailed description of our framework while highlighting the progress we have made since we published our initial results.
\label{sec:profilevisual}

\begin{figure*}[t]
\centering
\centering
\begin{small}
\includegraphics[width=13cm]{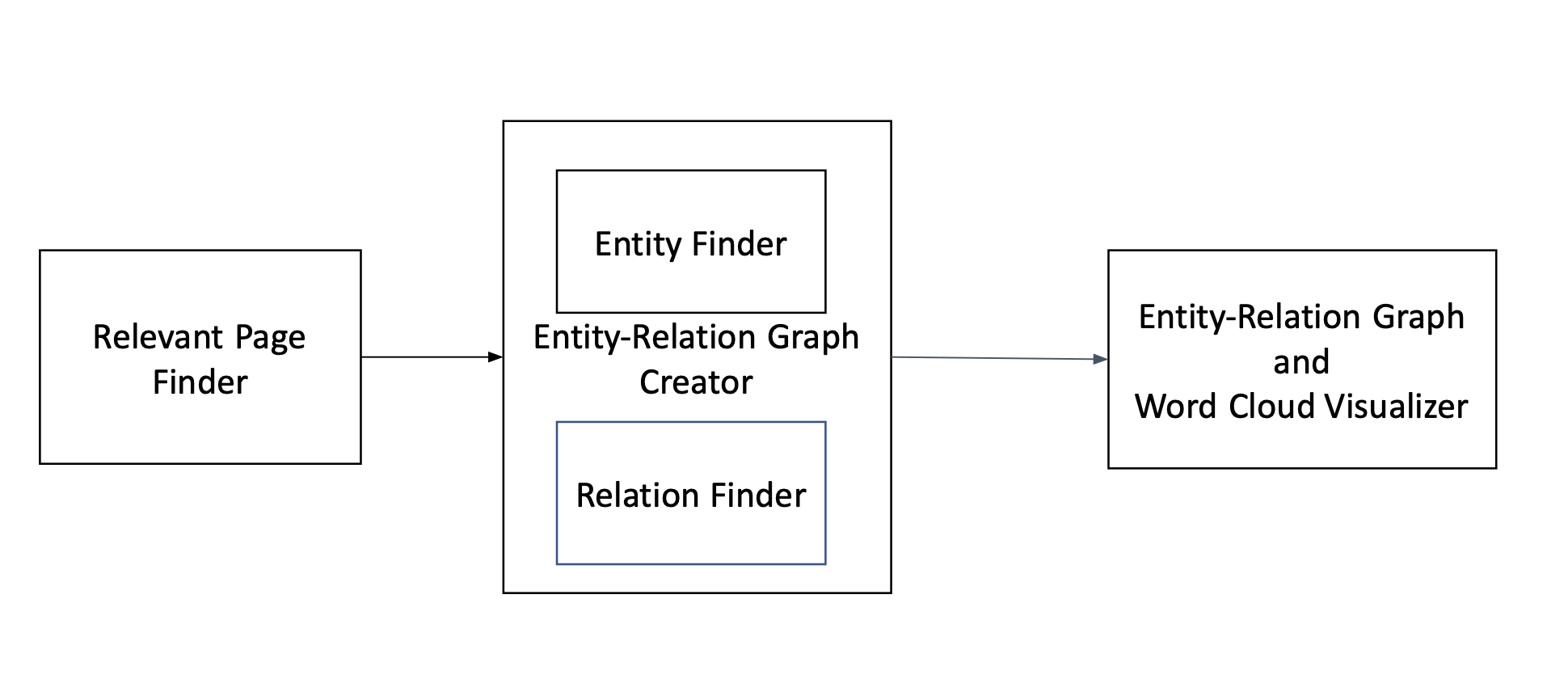} 
\caption{Flow chart illustrating the process of entity relation graph creation and visualization.}
\label{fig:flowchart}
 \end{small}
\end{figure*}

Our method is a generalized, comprehensive and larger scale method for the graph-based entity profiling method presented by \cite{amal2019relational} and includes the following steps. First, relevant pages are identified, in contrast to the aboved- mentioned work that focused on finding homepages of the entity in question. Then their content is analyzed by extracting from them the named entities and their relations to the entity in question. To extract of named entities, unlike the above-mentioned work, we performed massive evaluation on larger scale datasets and on a variety of domains, beyond academics that were used in previous studies. For the typed relation extraction, we proposed new techniques that involved deep learning and evaluated them together with the previously used techniques on larger scale datasets and on a larger variety of relation types. Finally, the constructed profile is visualized using a 2D graph and a word cloud, in opposition to the above-mentioned work where basic textual presentation was presented to the user. Figure ~\ref{fig:flowchart} shows the flow chart of this new process.

\subsection{Graph Profile Construction} 

Given regular query results (obtained from a regular search engine, Google in our case) about a named entity {\it t}, we identify highly relevant Web pages (relevant sources) in the results and then build a graph profile for that entity Gt, presenting its interconnections with related, typed entities {\it Et} over typed relations: \[ r(t;e); e \in Et, \] To automatically generate Gt from Web-based information, we follow the following process.

\subsubsection{Identifying relevant sources}

Given an entity name, specifically an entity name that denotes a person, our objective is to identify highly relevant Web pages that concern the entity, such as the person’s homepage, or their profile on a public social network, if available.
In addition, we look for other Web pages that may contain information about the entity such as interviews of the person, news articles about the person, mentions of the person in related organizations or by colleagues (if the entity is a person of interest), etc. Once identified, these Web pages can serve as the source for factual information about the entity. 
Given that in this research our focus is on persons, we identified a set of person entities from several domains ({\it scholars, actors, singers, sports people, politicians, financiers} and {\it lawyers}) and then queried Google, and obtained the top-k results. While the retrieved results were generally of high quality, we found that it is often the case that the results at the top of the retrieved list contained irrelevant pages, whereas the relevant results were actually located further down the list. This is to say, for a sample of 97 person entity names for whom Google retrieved 1.3K Web pages ~\footnote{https://github.com/saeedaml/relevant-page-finder-source-and-data/blob/main/README.md}, we found that 64 of the retrieved Web pages were not relevant and still appeared among Google’s top five results, and 230 were relevant but appeared between the ranks 10\textendash20 of the results retrieved by Google. Hence, we defined and implemented a supervised learning approach for the task of finding relevant pages. We manually labeled the constructed dataset of 1277 Web pages retrieved by Google as either relevant (786) or irrelevant (491) Web pages. Assuming that the search engine ranking is sensible yet imperfect, we modeled the original rank of the Web page in the retrieved initial list as a feature. Overall, we considered up to top 20 Web pages (some of the persons had less) retrieved by Google as candidate results in our experiments.  We constructed the following feature types for automatically assessing Web page relevancy. In total we included 1.3K Web pages, which consisted of the following:
\begin{itemize}

\item Diverse features: The majority were lexical features that denote the bag-of-words representation of the Web page. They included the Web page’s title, URL, body and the page's image.

\item Page type: Indicates whether the Web page originates from any of the following public social networking sites: Facebook, LinkedIn, Google Scholar, Twitter, ResearchGate, GooglePlus. 
\item Title relevance: Indicates whether the URL and/or the title of the Web page (based on the HTML markup) contains the entity name or its variants. In general, it is expected that personal homepages and other highly relevant Web pages have the person's name or its variants mentioned in these fields.
\item Title length: We denoted the number of words included in the Web page’s title as a feature.
\item Body (content) relevance: We counted the number of times the person's entity name or its variants is mentioned in the body of the Web page, as it is expected that in highly relevant Web pages, the person's name or its variants would appear multiple times in the page content. 
\item Images: we detect whether the page contains an image containing the person's entity name or its variants (within the caption or other relevant HTML markup).
\item Webpage length: We modeled the total number of words included in the body of the HTML for other features. As for the other features, the word count is discretized into Boolean features (indicating whether the number of words was less than 50, less than 100, less than 500, or greater than 500). This feature was motivated by the conjecture that homepages and other highly relevant Web pages are characterized by a typical page length.
\item Web page rank: Indicates Whether the Web page is ranked first in Google's list of Web pages.
\end{itemize}

Using the above-mentioned features, we built a vector representations for each of the Web pages. We then use them for training the classifier that is that later on is used to classify Web pages as relevant or not (as an information source about the entity). Once multiple classification models using the labeled examples are trained, given a new person entity (unseen by the models) the relevant Web sources about this specified person entity are obtained as follows. First, we searched Google using the name of the entity of interest as a query. Then, we applied the best performing classifier to evaluate the relevancy of each item in the top-ranked set of retrieved Web pages. We evaluated the performance using tenfold cross-validation in terms of precision, recall, and F1, as described in Section 4.1. 

\subsection{Fact Extraction}
 
Having identified relevant Web pages, we identify the named entities mentioned in these pages. We then wish to identify the relations that hold between the entity of interest and each of the identified entities, based on contextual evidence within the page. Once this information is obtained, it lends itself to creating a relational graph that describes the entity of interest. We now describe our approach for addressing these tasks.

\subsubsection{Named entity tagging}
We used several types of content for entity identification on the target Web pages. First, assuming that the hyperlinked text corresponds to related concepts and entities, we leveraged hyperlinks that are often available on a Web page ~\citep{blancoEOS11}. It is often the case, however, that entity names are merely mentioned in the text. Therefore, as a second step, we employed text processing tools to identify named entity mentions and their semantic types, including person, location, and organization. There exist numerous IE methods ~\citep{Sarawagi08} and tools that perform Named Entity Recognition (NER) in free text, such as URES ~\citep{Rosenfeld06}, DBpedia Spotlight ~\citep{mendes11}, TextRunner ~\citep{banko07}, KnowItAll informatics system ~\citep{Etzioni05} and the Stanford named entity tagger ~\citep{finkel05}. In our research, we adopted the latter tool, which is widely-used. Presumably, there exist many entities that are related to the target person and are not mentioned on their personal homepage. As already noted, we single out such entities by identifying and analyzing highly relevant – vis-a-vis the entity of interest – Web pages and not only the entity’s homepage. Specifically, we address the task of finding three kinds of related entities. The first type are those appearing on relevant pages, i.e., the personal Web pages of the entity of interest; the second type are those that do not appear on the personal Web pages nor on the entity’s personal Wikipedia page(s) and the third type are those that do not appear on the entity’s personal Web pages but do appear on the entity’s personal Wikipedia page(s). Since Wikipedia is well edited, the latter type has a validated capability of containing mentions of entities related to the entity of interest beyond the personal Web pages of the latter when they have no Wikipedia page.

\subsubsection{Relation type tagging}

The type of relation that holds between the person of interest {\it p} and each related entity {\it e} that was found in the previous step must be inferred based on the context in which {\it e} is mentioned on the relevant Web pages. To identify a relation between an identified entity and the entity of interest, we adapted pattern mining techniques, assuming that the relation type may be induced from the text that surrounds the related entity on the Web page  ~\citep{Fader11} ~\citep{Palkal05}. We chose to constrain the assigned relations to a set of predefined types so as to simplify the representation scheme, and support relational learning ~\citep{minkovTIST16}. Again, we addressed relation prediction as a supervised classification task: given the mention e of a named entity and local context represented as the surrounding words, the respective relation has to be predicted based on and from the set of the target semantic types, \[ r\in R \] The target types of interest may vary by the target domain. For example, in a study focused on scholars, the defined target relation types include the relations of education (linking the scholar with the name of the school that they attended, for example), publications (e.g., linking the scholar with the names of co-authors), and work (linking with organizations and peers in the context of work), etc.
We approached the task of relation type tagging as a classification task. As described in Section 4.2.2, we obtained relevant labels in a lightly supervised fashion, using distant supervised learning. As features, we considered the related entity name string and the words surrounding the relevant mention of the named entity (five words before and following the appearance of the related entity), including their distance from the related entity.

\subsection{Entity Profile Visualization}
\begin{figure*}[t]
\centering
\centering
\begin{small}
\includegraphics[width=13cm]{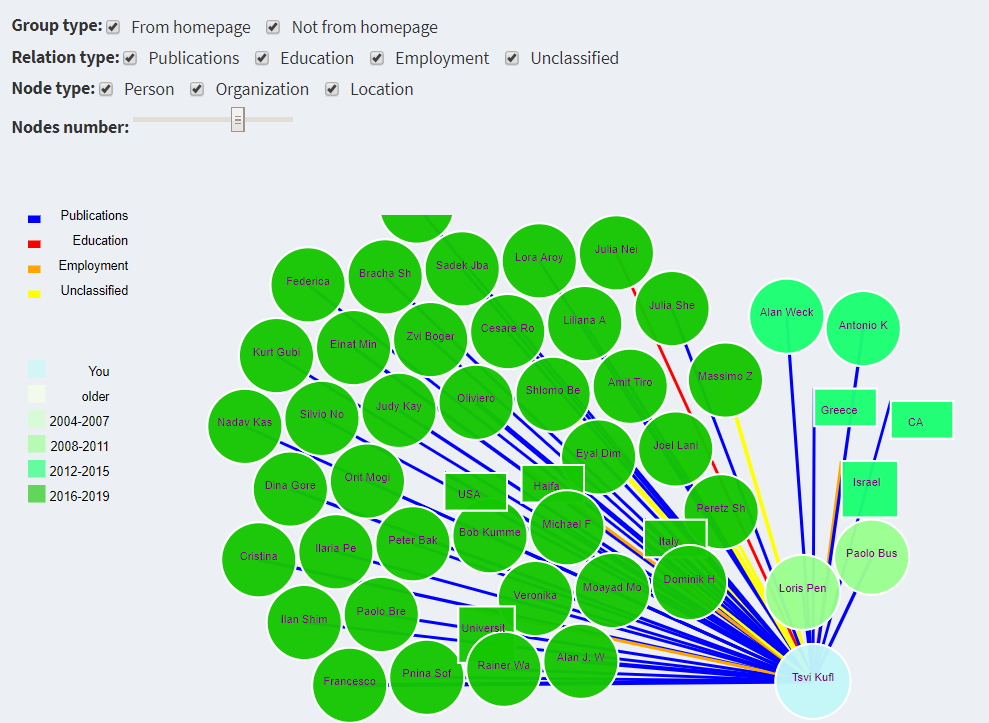} 
\caption{Example of our generated profile visualized using a 2D graph.}
\label{fig:prfileviz}
\end{small}
\end{figure*}

To enable easy examination of the graph-based profile, we built an interactive visual that displays it.
Notably, the number of related entities for a given entity of interest can be large. Displaying detailed information about the related entity types, the types of the relations with those entities, the strength of these relations and their temporal aspect may be overwhelming for browsing purposes. To accommodate and serve such rich data effectively to the user, we, therefore, visually differentiate between multiple facets, and support filtering out information that is considered extraneous by the user or enabling the focusing on specific subsets of the information. That is, we aimed for a graph profile that is intuitive, self-explanatory and interactive, and offers information at different levels of detail.
An initial version of the generated graph scheme visualizing the constructed profile of a scholar is illustrated in Figure~\ref{fig:prfileviz}. 
As shown, the various components of the structured profile are displayed using dedicated visualization techniques and are complemented using word cloud visualization. Specifically,
\begin{itemize}

\item The shape of the nodes denotes their type (circle for {\it Person}, square for {\it Organization} and rectangle for {\it Location}).
\item The color of the edges denotes the relation type (blue for joint {\it Publications}, red for {\it Education}, orange for {\it Employment}  and yellow for other relation types.
\item The width of the edges denotes the strength of the relation between the two entities, which is inferred from the number of times that the related entity was mentioned on the analyzed Web pages.
\item The color of the nodes denotes the temporal aspect. As mentioned before, we identified year expression in the surrounding context of the related entities. In this case we split the temporal aspect into five intervals: 0--2004, 2004-–2007, 2008-–2011, 2012-–2015, 2016-–2019. Specifically, the color ranges from light green for related entities in the farther past, to darker green for entities mentioned more recently, based on the most recent period associated with the relation with that entity. When no year information is found, the node is white. Thus, the user can focus on specific temporal intervals, e.g., focus on persons that related to them more recently, provided this information is available.
\end{itemize}

\begin{figure*}[t]
\centering
\centering
\begin{small}
\includegraphics[width=10cm]{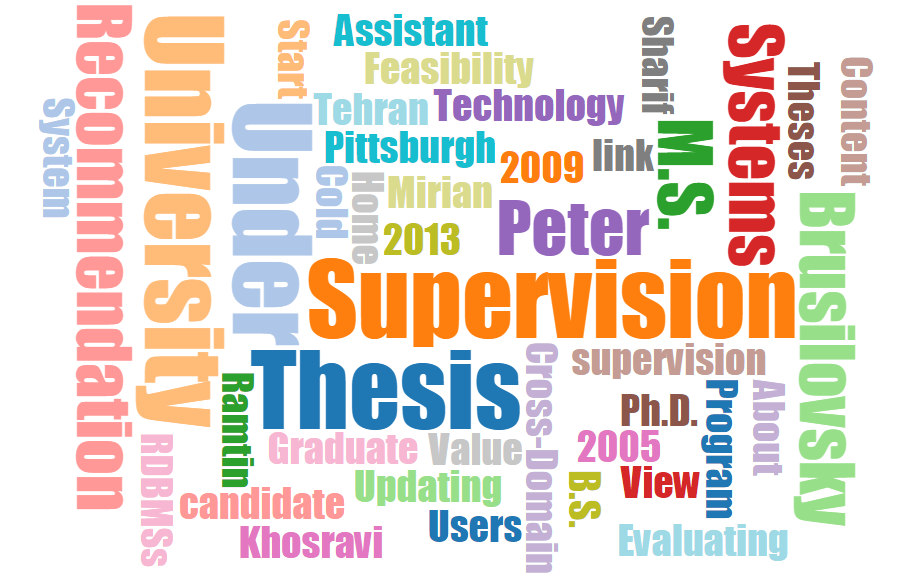} 
\caption{Example of our word cloud shown on an edge in the graph.}
\label{fig:wordcloud}
 \end{small}
\end{figure*}

For enhanced explainability, once the user clicks on an edge, a word cloud pops up (see example in Figure~\ref{fig:wordcloud}), presenting the text surrounding the related entity mention(s) in the form of a word cloud.

As shown in Figure~\ref{fig:prfileviz}, the user has the option of exploring the visualized graph using multiple filters, available as selectable checkboxes. These filters pertain to:

\begin{itemize}
\item Underlying information source: Homepage vs. other relevant pages 
\item Relation type: Displaying entities linked over a specific edge type (Publications, Education, Employment, or other)
\item Type of the related entities: Person, Organization or Location.
\end{itemize}
An additional supporting swipe bar allows the user to control the number of nodes displayed. This functionality enables the user to focus their attention on the top weighted 10/20/../70/all related entities that match the specified criteria.  
Implementation-wise, we considered the Pajek, GraphViz and Gephi tools/APIs, which are often used for ontologies or graph representation, but found them to be slow in rendering and interacting with dense graphs. Hence, we opted to use the D3.js data-driven document library and its implementation of forced directed graphs, also known as spring embedders. Such algorithms calculate the layout of a graph using only information contained within the structure of the graph itself, rather than relying on domain-specific knowledge ~\citep{Kobourov12}. The graph is plotted as a treemap, a method for the visualization of hierarchically structured information. It uses 100\% of the available display space, mapping the full hierarchy onto a rectangular region in a space-filling manner ~\citep{johnson91}, where the root is the logged-in user (the target entity) and the rest of the nodes are the related entities. We found this tool to be very useful: it is dynamic, scalable and interactive. The graph profiles were visualized using our implemented layer on top of the implementation of the forced directed graph via the D3.js data-driven document library.

\section{Experiments: Evaluation of the Framework's Components}
\label{sec:experiments}

To demonstrate and evaluate the suggested framework, we implemented it and then conducted a series of studies that assessed the individual system components one by one. Following this, an end-to-end evaluation was also performed. Specifically, we evaluated: 
 \begin{itemize}
\item Finding relevant pages  
\item Profile graph construction, comprising the tasks of related named entity extraction and tagging of the relations with those related entities
\item Entity-Relation Graph, Word Cloud Visualizer and End-to-End Framework
\end{itemize}

In what follows we describe the various studies. For each study, we describe the dataset, experimental that was used, the experimental procedure and the results.

\subsection{Automatic Identification of Relevant Pages}

First, we explored the potential of identifying relevant Web pages that describe an entity of interest, focusing on {\it person} entities. In earlier work as in ~\cite{amal2019relational}, the focus was on automatic identification of the entity's homepages on a small dataset. Here we go beyond and retrieve any highly relevant Web pages and also evaluate our system on a larger number of entities and Web pages. We also consider and evaluate how deep learning helps to improve the results. Having retrieved a ranked list of Web pages given the person’s name, we automatically assess the relevance of these pages. 

Our experimental study comprises two parts: first, we fine-tune a classifier using a manually annotated dataset, which includes a diverse sample of target persons. Then, we scale up the learning process by using the trained classifier to automatically label a larger dataset of a particular population (congressmen) and then using the new corpus, we re-train the classification models for that population. We show that the learning, in general, is effective, yielding high-quality results, and that scaling up the process using the automatically-assigned labels can further improve the results

\subsubsection{Datasets}

We constructed several datasets to evaluate the page finder component. One dataset was small and diverse for evaluating a general task vis-à-vis any person type entity.  A second larger-scale dataset focused on a specific domain, which enabled us to evaluate how a deep learning model would perform on a specific task in a named domain (since deep learning requires a large amount of data for performance optimization and as we saw that the model did not perform well on the smaller set). Finally, we built a small test set based on the specific domain to evaluate how well the model performs.

The first dataset consists of Web-pages related to of 97 person entities from several domains, including scholars (32), actors (15), singers (5), sportsmen (5), politicians (10), financiers (20) and lawyers (10). The entities were selected randomly from the Web by searching for entities in each of the above-mentioned domains. The domains themselves were selected according based on where persons from the domain have presence on the Web. Examples of the queries that were used for picking names in these domains are: “Sportsmen in the US”, “Politicians in the US”, etc. We used Google to search for entity name and considered the top twenty retrieved results (if available) as possibly relevant. The Web pages were manually assessed for relevance. This resulted in a dataset of 1277 annotated Web-pages in total, comprising 786 relevant Web pages and 491 irrelevant ones. On average, this dataset includes 8.10 relevant pages (e.g., homepages, social network profiles, articles about the person, interviews with the person etc.) per entity (STD=1.56) and 5.06 irrelevant ones (STD=1.03).

We found that the list of the retrieved results was noisy, with irrelevant pages appearing near the top, and relevant results farther down the list. Specifically, there were, on average, 5\% irrelevant Web pages among the top five retrieved results and 15\% irrelevant pages among the top ten results. On the other hand, we found 18\% of the lower ranked pages (number 10\textendash20 on the list) to be relevant.

We further constructed a larger dataset focused on Politicians as follows. We randomly selected the names of 400 US congress members as of 2018. We queried Google using each name and considered the top 50 ranked results per person name. Each retrieved Web page was then labeled {\it automatically} using the best classifiers trained on the manually annotated dataset described above. As a result, the Congress member dataset included about 8418 Web pages assessed to be relevant, and about 4870 irrelevant  Web pages. Google retrieved less than 50 Web pages for some politicians. On average, 32 Web pages (STD=1.83) were retrieved per person, out of which 21 (STD=1.31) were classified as relevant, and 11 (STD=1.96) were classified irrelevant. Finally, we built a small distinct test set of twenty names of different congress members (not included in the previous dataset), for which we retrieved 760 Web pages in total (38 Web pages per name, STD=2.92), which were manually assessed for relevance. 

\subsubsection{Experimental Setup}
We experimented with the following classification methods:
\begin{itemize}

\item A deep neural architecture network that computes the representation of input Web page raw data using the encoder, considers all hidden states of the encoder using the attention mechanism and then predicts the class as Relevant or Irrelevant, using the decoder with the softmax classifier layer. We used a dropout rate of 0.5 and early stopping, and experimented with varying numbers of hidden layers (1, 5, 10, 15, 20, 50).  
\item Classification methods that are generally known to perform well on text classification problems, including SVM ~\citep{Joachims98}, C4.5 decision tree (J48 classifier) and REPTree (Reduced Error Pruning tree) decision trees ~\citep{Quinlan86}~\citep{yang94}, Naive Bayes ~\citep{kim06}, Bayesian Logistic Regression ~\citep{genkin07} , and K-nearest neighbors ~\citep{yang94}.
\item Ensemble classification approaches, including Random Forest ~\citep{Xu12}, Boosting ~\citep{Schapire2000}, Bagging ~\citep{liu03}, the Random Subspace method ~\citep{gangeh10}, and LogitBoost ~\citep{friedman00}.
\end{itemize}

For evaluation purposes, we performed  10-fold cross-validation using the annotated dataset, which is relatively small. The features used are described in Section 3.1. While representing the textual content of the Web pages as features, we de-lexicalized the names of the  entities of interest (replaced the name with a generic ‘x-person’ placeholder).

\subsubsection{Experimental Results}

Table~\ref{tab:webpageclassif} presents the results of the different classifiers on the manually annotated data in terms of precision, recall, and F1. For the deep learning architectures, the error measurement commonly used is the total cost of the loss function — specifically, in our experiments, cross-entropy.

\begin{table}[t]
\begin{center}
\begin{tiny}
\begin{tabular} {lllllll}
\hline
      {\it\bf Classifier} & {\it\bf J48} & {\it\bf NB} &
      {\it\bf SMO} & {\it\bf  KNN} & {\it\bf BLR} & {\it\bf PEPT} \\
      
\hline
{\it\bf F1} & 0.79 & 0.75 &{\bf 0.8} & 0.74 & {\bf 0.8} & 0.77 \\

\hline
{\it\bf Precision} & 0.79 & 0.8 & 0.78 & 0.8 & 0.8 & 0.78\\
\hline
{\it\bf Recall} & 0.79 & 0.71 & 0.82 & 0.69 & 0.81 & 0.75\\
\hline
\hline 
      {\it\bf Ensemble} & {\it\bf Bag.} & {\it\bf Boost.} & {\it\bf R.Forest} & {\it\bf R.Sspace} & {\it\bf L.Boost} \\
\hline
{\it\bf F1} & {\bf 0.92} & 0.8 & 0.78 & 0.82 & 0.82\\

\hline
{\it\bf Precision} & {\bf 0.9} & 0.75 & 0.78 & 0.79 & 0.76 \\
\hline
{\it\bf Recall} & {\bf 0.93} & 0.85 & 0.79 & 0.85 & 0.88\\
\hline
\hline 

      {\it\bf Deep Learning} & {\it\bf DNN-1} & {\it\bf DNN-5 DO=0.5} & {\it\bf DNN-10 DO=0.5 } & {\it\bf DNN-15 DO=0.5} & {\it\bf DNN-20 DO=0.5}& {\it\bf  DNN-50 DO=0.5} \\
\hline
{\it\bf F1} & 0.77 &0.71 &0.71 &0.69 &0.73 &0.76 \\

\hline
{\it\bf Precision} & 0.76 &0.70 &0.70 &0.69 &0.74 &0.74\\
\hline
{\it\bf Recall} & 0.78 &0.72 &0.72 &0.68 &0.71& 0.77\\
\hline
{\it\bf Loss cost} & 1.05 &1.09& 1.09& 1.12& 1.07& 1.05\\
\hline

      {\it\bf Deep Learning } & {\it\bf DNN-1 } & {\it\bf DNN-5 ES } & {\it\bf DNN-10 ES } & {\it\bf DNN-15 ES } & {\it\bf DNN-20  ES }& {\it\bf  DNN-50 ES} \\
\hline
{\it\bf F1} & 0.74 &0.8 &0.81 &0.81 &0.81 &0.8 \\

\hline
{\it\bf Precision} & 0.75 &0.78 &0.79 &0.8 &0.79& 0.78\\
\hline
{\it\bf Recall} & 0.73 &0.82 &0.82& 0.82& 0.82& 0.81\\
\hline
{\it\bf Loss cost} & 1.07 &0.65 &0.58 &0.52 &0.57 &0.62\\
\hline
\hline

\end{tabular}
\end{tiny}
\end{center}
\caption{\label{tab:webpageclassif}  Experiment 1 – Statistics of the different classifiers. Bold fonts in the F1 rows indicate statistical significance (Kruskal-Wallis, p \textless 0.05)\\
*DO=Drop Out *ES=Early Stopping *DNN=Deep Neural Networ}
\end{table}

As shown in Table~\ref{tab:webpageclassif}, the bagging approach achieved an F1 of 0.92, outperforming the other classifiers. The deep neural architecture with early stopping performed than the dropout network; specifically, the architectures that had 5 and up hidden layers yielded an F1 of 0.8\textendash0.81, the second-best result. The lowest results were obtained using the deep neural architecture with a dropout rate of 0.5. Also, we can see that for the deep neural architecture, as the number of hidden layers grows, the F1 does not improve and even starts decreasing, probably because of overfitting. As expected, the deep learning architecture performed relatively poorly because of the small training set, which may be too small for training a deep learning model. 

\paragraph{Feature analysis}

Overall, our feature set included 1.3K attributes as described in Section 3.1.1.
We measured the information gain with respect to the target classes using Weka ~\citep{Frank16}. The top five attributes with the highest information gain ratio are detailed in Table~\ref{tab:IG_attrib}. The most significant feature denotes whether variants of the person’s name appear in the body of the Web page and the next two features indicate whether the exact full name of the person appears in the body and the title, respectively. Other features with high information gain include the mentions of words such as ‘university’, ‘finance’, and ‘research’ in the body of the text. It is interesting to note that at rank 34 (when ranking features in descending order using the information gain scores of the features), we see the attribute of whether this Web page is ranked first in Google’s list of Web pages.

\begin{table}[t]
\begin{center}
\begin{small}
\begin{tabular} {ll}
\hline
        {\it\bf Feature name} & {\it\bf IG}\\
\hline
       {\it\bf Variants of the person's name in the body of the HTML} & 0.05\\
\hline
       {\it\bf Exact full name of the person appears in the HTML  }& 0.047\\
\hline
       {\it\bf  Exact full name of the person in the title of the HTML } & 0.047\\
\hline
       {\it\bf Variants of the person's name in the URL } & 0.046\\
\hline

      {\it\bf Variants of the person's name appear in the title of the HTML}  & 0.045\\
      
\hline
\end{tabular}
\end{small}
\end{center}
\caption{\label{tab:IG_attrib}  Information gain (IG). Most significant attributes ranked in descending order}
\end{table}

\paragraph{Scaling Up}
 Next, we focused on the thus far best-performing classifier - namely, Bagging, and the best performing deep learning configuration. We applied the classification models learned using the manually annotated dataset to automatically label our larger dataset of Politicians.
 We first computed recall with respect to the set of Web pages that were retrieved per query entity (for details about the congress member dataset, see Section 4.4.1) by the search engine, which we manually labeled as relevant. The performance in terms of F1 increased slightly compared with our experiments on the smaller dataset, increasing from 0.92 to 0.93 (Precision\textasciitilde0.92,Recall\textasciitilde0.94), despite the automatic label setting (which is expected to be less accurate). This is likely due to the increased dataset size. As for the deep learning architecture, training on this larger dataset improved performance substantially. It outperformed Bagging, reaching F1\textasciitilde0.98 (Precision\textasciitilde0.97,Recall \textasciitilde 0.98).

\begin{table}[t]
\begin{center}
\begin{tiny}
\begin{tabular} {llllllll}
\hline
      {\it\bf List length } & {\it\bf STD  }& {\it\bf Precision - Bagging } & {\it\bf Recall - Bagging} &
      {\it\bf  F1 – Bagging} & {\it\bf  Precision - DL  } & {\it\bf Recall - DL } &
      {\it\bf F1 - DL} \\
      
\hline
38 & 2.92 & 0.92 & 0.94 & 0.93  & 0.97 & 0.98 & 0.98 \\
\hline
\end{tabular}
\end{tiny}
\end{center}
\caption{\label{tab:testcongress}  Classification results for 20 test congress members, using a training set labeled automatically.}
\end{table}

In summary, we conducted two experiments for evaluating the ability to automatically identify relevant Web pages about an entity of interest while initially retrieving candidate Web pages using Google. Our experiments yielded high quality results over different domains, both at small and larger scales.

\subsection{Entity-Relation Graph Creation}
We next describe two studies, intended to assess the extent to which our framework can identify relevant related entities, and the relations with these entities, using the Web pages identified as relevant for an entity of interest. In terms of relation extraction, we are interested in correctly identifying the relation types. We first consider salient relations within the scholar domain and then extend this study to a more varied set of general relation types.

\subsubsection{Entity Finding}
In this study, we attempted to assess the ratio of related entities that do not appear on a person’s homepage. In particular, we addressed the question of how often are related entities (as a reference we used those appearing on the entity of interest’s Wikipedia page) not mentioned on the entity of interest’s personal homepage? This study involved the construction of dedicated datasets, as described below. While in previous work as ~\cite{amal2019relational} the Stanford Named Entity tagger~\citep{finkel05} was also used to identify related entities, here as mentioned above we focus mainly on the evaluation of the full process up to this point while considering all relevant Web pages and not only homepages which introduce the challenge of noisy data. 

\paragraph{Datasets}

To construct the dataset, we considered 6750 person names from four different domains: politicians (1800), businessmen (950), actors (3150) and singers (850). We queried Google with their full names and considered up to fifty top-ranked Web pages for each person. The average number of Web pages obtained was typically low and differed per population: politicians (34), businessmen (28), actors (39), and singers (46). We used the Bagging classifier (not the Deep Neural Network (DNN) since it was evaluated on the specific domain of {\it politicians}) described in the previous study (Section 4.1) to classify these Web pages as relevant or irrelevant to the entity of interest. 
In Table ~\ref{tab:statfourdomains}, we detail the statistics for this set of entities. In total, we extracted 886537 entities for whole set of 6750 person names.

\begin{table}[t]
\begin{center}
\begin{tiny}
\begin{tabular} {llllllllllll}
\hline
{} &{\it\bf Total }& {\it\bf STD }  & {\it\bf \#E} & {\it\bf   \makecell{\#NH \\ (All)}} & {\it\bf  \makecell{\#NH \\ (Per)}} & {\it\bf  \makecell{\#NH \\ (Org)}} & {\it\bf  \makecell{\#NH \\ (Loc)}} & {\it\bf  \makecell{\#NHW \\ (All)}} & {\it\bf  \makecell{\#NHW \\ (Per)}} & {\it\bf \makecell{\#NHW \\ (Org)}} & {\it\bf  \makecell{\#NHW \\ (Loc)}} \\

\hline
{\it\bf Politicians} &237946 &25 &127 & \makecell{36\\ (28\%)} & \makecell{18 \\(14\%)} & \makecell{11 \\(9\%)} & \makecell{7 \\(5\%)}& \makecell{4 \\(11.5\%)} & \makecell{1 \\(3\%)}& \makecell{1 \\ (3\%)} & \makecell{2 \\(5.5\%)}\\

\hline
{\it\bf Businessmen} &108335 &22 &111 & \makecell{32\\ (29\%)} & \makecell{17 \\(15.5\%)} & \makecell{9 \\(8\%)} & \makecell{6 \\(5.5\%)}& \makecell{3 \\(9\%)} & \makecell{1 \\(3\%)}& \makecell{1 \\ (3\%)} & \makecell{1 \\(3\%)}\\

\hline
{\it\bf Singers } &109608 &33 & 128 & \makecell{42\\ (33.5\%)} & \makecell{24 \\(19\%)} & \makecell{11\\ (9\%)} & \makecell{7 \\(5.5\%)}& \makecell{6 \\(14.5\%)} & \makecell{3 \\(7\%)}& \makecell{1 \\(2.5\%)} & \makecell{2 \\(5\%)}\\

\hline
{\it\bf Actors } &430648 &31 & 143 & \makecell{34\\ (24\%)} & \makecell{16 \\(11\%)} & \makecell{12\\ (8\%)} & \makecell{7 \\(5\%)}& \makecell{7 \\(21\%)} & \makecell{3\\ (9\%)}& \makecell{1\\ (3\%)} & \makecell{3 \\(9\%)}\\
\hline

\end{tabular}
\end{tiny}
\end{center}

\caption{\label{tab:statfourdomains}  Statistics on contribution from four different domains: Average number of related entities found per person (E); average number and ratio of related entities not found on the entity’s homepage (\#NH); and average number and ratio of the total related entities, which were not found on the entity’s homepage but are mentioned on the entity’s Wikipedia (\#NHW).}

\end{table}
‬‬‬‬‬‬‬‬‬‬‬‬‬‬‬‬‬‬‬‬‬‬
\paragraph{Procedure}
We wished to gauge the added value of identifying relevant related entities on Web pages beyond the entity of interest’s homepage. We first identified named entities on the relevant Web pages retrieved per entity of interest using the Stanford Named Entity tagger~\citep{finkel05}, then we checked whether each of these named entities appeared on the entity of interest’s homepage. To validate the correctness of the additional extracted entities, we checked whether they appear on the entity’s Wikipedia page. We evaluated this aspect in four different domains (politicians, businessmen, actors, singers).

\paragraph{Results}
The results are detailed in Table~\ref{tab:statfourdomains}. We can see that across all domains, we were able to find relevant entities (we checked whether those entities were mentioned on the entity of interest’s Wikipedia page, as an indication of their relevance and importance) that were not included on the entity of interest’s homepage. We observe that there is a high number of entities that entities of interest do not mention on their homepages (\#NH). Actors have the lowest percentage of new entities (24\%) and singers have the highest (33.5\%). We note that the majority of these nonmentioned entities are of the person type, followed by organizations. As detailed in Table ~\ref{tab:statfourdomains} (in the columns titled \#NHYW), the ratio of named entities found that do not appear on the entity of interest’s homepage was highest for actors (21\% out of 24\% overall), and lowest for businessmen (9\%). It is further shown that the majority of those entities are of the person and location types. This may indicate that people tend to mention their relationships with organizations more than with people.

The following are several examples, which suggest that named entities that are not mentioned on one’s homepage often reveal interesting facts.

\begin{itemize}

\item {\bf[Business people: Christina Lewis]} - This entity of interest appeared with the entity Reginald Lewis in the context of “Reginald Lewis’ Christina Lewis opens up about growing up with her famous father”. While her father is mentioned on her Wikipedia page, he is not mentioned on her homepage. 

\item {\bf [Business people: Cristina Patwa]} – Patwa was co-mentioned with Robert Rodriguez on a relevant Web page and on her Wikipedia page, but not on her homepage: “In December 2013 Patwa partnered with Robert Rodriguez to create the El Rey network, an English language American television channel targeting Latino audiences”.

\item {\bf [Politicians: John Robert Beyster]} – Interestingly, Beyster does not mention on his homepage his relationship with the University of Michigan, where a relevant Web page stated that on Wednesday April 11 2012, the computer science engineering building at the University of Michigan was rededicated in his name as the Bob and Betty Beyster Building.

\item {\bf [Actors: Oprah Winfrey]} – Facts of interest found on a relevant Web page that do not appear on her homepage refer to her relationship with the entity Vernon Winfrey, her father: “Winfrey was then sent to live with the man she calls her father, a barber in Tennessee and landed a job in radio while still in high school".
\end{itemize}

In summary, in these cases, we were able to find related entities on other relevant Web pages that were not mentioned on the entity of interest’s homepage. While we were able to validate the relevance and correctness of these entities by considering the Wikipedia pages of the entity of interest as our gold standard, many entities do not have a Wikipedia page. Our approach, which considers multiple relevant pages as sources of information about the entity of interest, is, therefore, expected to yield richer and more complete factual descriptions of such entities. A limitation of our evaluation, however, is that we do not report the false positives, the entries that are found neither on the person’s homepage nor on their Wikipedia page because of the large number of entities that need to be validated. For this reason, we report the above few examples.

\subsubsection{Relation Finding}
Next, we evaluate the possibility to automatically assign a semantic type, out of a pre-specified set of relation types, to a relation between a target person p and each related entity e found on the related Web pages. The evaluation is done in two case studies, in different domains, including a scholars dataset and a mixed dataset that includes politicians, businessmen, actors and singers. Whereas in previous work as ~\cite{amal2019relational}, the focus was on a small set of data, here we focus mainly on the construction of large-scale datasets and the evaluation of the full process up to this point considering all relevant Web pages, which introduce the challenge of noisy data, and not only homepages.

\paragraph{Datasets}
\subparagraph{Scholar domain}
In the scholar domain, we applied distant supervision ~\citep{mintzACL09} to label relation types. Assuming that the scholar’s CV is available, one may determine the relation type with an entity mentioned in the CV by nothing the section in which it is mentioned. We discuss this proposed distant supervision labeling scheme below. Accordingly, we constructed a list of 300 scholars that we selected randomly from various universities and departments around the world, for whom we were able to access both their homepage and CV. For those academics, we extracted 1374 unique named entities from the homepages that also appeared in their CVs and whose type we were able to define  (the section name defines the type).

The labeling process is as follows. To avoid costly manual annotation of a large number of relation instances, we devised a domain-specific distant supervision labeling scheme. Let the target set of relation types be defined, without loss of generality, as \[R = \{{\it education, employment, publications, other}\} \] Leveraging the fact that {\it scholars} often make their CV available on their Web page, we sought homepage and CV pairs. Having extracted named entities from the homepage, we then sought those entities in the CV. Once we mapped the section titles to the target set of relation types R, we automatically determined the relation types based on the CV sections in which the entity was mentioned. For example, let us assume that the entity 'University of Zurich' was detected on the respective scholar’s homepage, as well as in their CV in a section titled 'Academic Education'. Using a set of manually-crafted mapping rules, the relation was mapped in this case to the generic relation type {\it education}. Likewise, co-author names and venues, which appeared in CV sections describing 'publication history', were assigned the relation type {\it publications}. The resulting automatically-annotated dataset included 605 entity names assigned the relation type {\it employment}; 581 entities annotated as {\it publications}; and 188 entity names annotated as education.

\subparagraph{Multiple domains}
To assess the generality of our framework, we constructed another dataset, in which the target persons comprised {\it politicians, businessmen, actors,} and {\it singers}. Our dataset included focus entities, related entities and the relations with their interrelations, extracted from an NYT10 dataset prepared by Riedel~\footnote{http://iesl.cs.umass.edu/riedel/ecml/ in (Sebastian Riedel, Limin Yao, and Andrew McCallum. Modeling relations and their mentions without labeled text.)}. They employed distant supervision to perform relation extraction, where the data consist of automatically labeled relational information. Overall, the resulting dataset consisted of 500K inter-entity relations and their textual context, mapping to 52 relation types in total ; for example: the capital of a country, the place where the company was founded, a person’s place of birth or ethnicity, nationality, places lived during their professional life, relations related to the religion of a person, geographic information etc.

\paragraph{Procedure}
We encoded the features for classifying the relation with each related entity into its semantic type: using the bag of words approach that includes the name of the related entity of interest,  as well as five words before and five words after the related entity name. Additionally, we used positional features, which indicate the word information as well as its distance from the related entity.

For the {\it scholar} dataset, which is relatively modest in size, we performed 10-fold cross-validation, using the classifiers detailed in Section 3.2, where for each target relation type, we trained dedicated binary classifiers as J48 decision tree, logistic regression (LR), SVM and Naive Bayes (NB) and evaluated them using a balanced set of positive examples and negative examples.

For the larger and diverse dataset, we applied a following network where the first input layer receives the word embedding representation using Word2Vec ~\footnote{https://code.google.com/p/word2vec/} and the second layer is a piecewise convolutional neural network (PCNN) fed by the previous layer. The third layer implements an attention mechanism fed by the PCNN output from the previous layer and the fourth layer is a softmax layer for classification.
Since the dataset was large, we trained the model using 80\% of the training examples and assessed performance on the remaining 20\% examples, used as a test set.

\paragraph{Results}
Table ~\ref{tab:reltypeclassif_2} reports the experimental results for the smaller dataset of scholars, detailed per target relation type and for the different evaluated classifiers. As shown, the best F1 score was achieved using a logistic regression classifier, with an average F1 of 0.91 for education and employment, while the highest F1 in the publication domain was the SVM with an F1 of 0.93). Since the logistic regression classifier achieved the above noted high average F1 across all categories, we used it in the second study to automatically identify the relation types of extracted named entities based on the context. As can be seen in Table ~\ref{tab:reltypeclassif_2}, we marked in bold the highest values across each category and evaluation metrics. For each of the three categories where we report the statistical significance, we used the Kruskal-Wallis Test and found that the results (F1s marked in bold) are statistically significant with p\textless0.05.
\\

\begin{table}[t]
\begin{center}
\begin{tiny}
\begin{tabular} {llllllllllllll}
\hline
      {\it\bf Relation type }  & {\it\bf \#examples} & {\it\bf  P-NB} &   {\it\bf  R-NB}&   {\it\bf  F1-NB}
      & {\it\bf  P-J48} &   {\it\bf  R-J48}&   {\it\bf  F1-J48}& {\it\bf  P-SVM} &   {\it\bf  R-SVM}&   {\it\bf  F1-SVM}& {\it\bf  P-LR} &   {\it\bf  R-LR}&   {\it\bf  F1-LR}\\
      
\hline
{\it\bf Publications } &   581 & 0.83 & 0.89 & 0.86 & 0.89 &0.81& 0.85& {\bf0.93}& 0.9 &{\bf0.92}& 0.91& {\bf0.91}& {\bf0.91}\\

\hline
{\it\bf Education } & 188  & 0.9 & 0.8 & 0.85 & 0.84 &0.86 &0.85 & {\bf 0.94} &0.81 &{\bf0.87}& {\bf 0.94}& {\bf0.92}& {\bf0.93}\\

\hline
{\it\bf Employment } & 601 & 0.88 & 0.77 & 0.82 & 0.74 &0.76 &0.75& {\bf0.89} &0.78& {\bf0.83}& 0.88& {\bf0.87}& {\bf0.87}\\
\hline

\end{tabular}
\end{tiny}
\end{center}
\caption{\label{tab:reltypeclassif_2} Summary of the evaluation of the relation type classification task in the scholar domain. Bolded numbers in the F1 column indicate statistical significance (Kruskal-Wallis, p\textless0.05) per relation type.
*P=Precision; R=Recall}

\end{table}

The statistics on the larger dataset are detailed in Table ~\ref{tab:riedeltypes}. The performance on the test examples, averaged over all the target 32 classes/relation types, achieved a precision of 0.73 and recall of 0.69 (an F1 of 0.71). Considering the large number of target classes, it is possible that while the dataset is large (400K), increasing the dataset would improve results further. 

We used this trained model to classify the relation types found on the dataset of relations that we created from the four domains: politicians (250802 tuples of relations), businessmen (112594), actors (480326), and singers (111989). Table ~\ref{tab:riedeltypes} details the statistics of classified relation types for each domain. As shown, a dominant relation was {\bf location} (e.g., the capital of a country, the place where the company was founded, or a person’s place of birth). Other common relations {\it include {\it ethnicity}, nationality of a group, and places lived} during their professional life, and {\it advisors} of company. On the other hand, the least common relations are the person's religion, geographic distribution, the location of a sports team, and relations related to shopping centers.

\begin{table}[t]
\begin{center}
\begin{tiny}
\begin{tabular} {lllll}
\hline
      {\it\bf Category name } & {\it\bf Politicians} & {\it\bf Businessmen} &
      {\it\bf Actors} & {\it\bf Singers} \\
      
\hline
{\it\bf Total number of examples tested per domain} & 250802 & 112594 & 480326 & 111989 \\
\hline
{\it\bf location - administrative division or country} & 54 & 20 & 37 & 9 \\
\hline
{\it\bf place of burial of person} & 4702 & 1775 & 30872 & 7653\\
\hline
{\it\bf featured film locations} & 91 & 30 & 339 & 110 \\
\hline
{\it\bf location – neighbourhood of} & 8239 & 4019 & 10382 & 2785 \\
\hline
{\it\bf location - administrative capital in an Indian state } & 88 & 45 & 188 &  56 \\
\hline
{\it\bf people with a profession}& 17 & 3  & 34 &  12 \\
\hline
{\it\bf location – capital in a Brazil state}& 60 & 30 & 367 & 102 \\
\hline
{\it\bf profession of person} & 57 & 28 & 199  & 52 \\
\hline
{\it\bf family's country} & 2 & 2 & 16 & 9 \\
\hline
{\it\bf contains location}& 990 & 538 & 1401 & 290 \\
\hline
{\it\bf locations of company }& 286 & 140 & 703 & 249 \\
\hline
{\it\bf children }& 346 & 341 & 699 &  197 \\
\hline
{\it\bf major shareholders in company}& 47 & 24 & 254 & 74 \\
\hline
{\it\bf founders of a company}& 2528 & 1541 & 2383 & 560 \\
\hline
{\it\bf place of birth of person}& 571 & 398 & 1019 & 557 \\
\hline
{\it\bf location - legislative capital in Indian state} & 0 & 0 & 0 & 0 \\
\hline
{\it\bf ethnicity of a group}& 24510 & 8408 & 49913 & 11348 \\
\hline
{\it\bf location - administrative divisions or countries} & 4 & 1 & 4 & 4 \\
\hline
{\it\bf religion} & 1 & 2 & 3 & 0 \\
\hline
{\it\bf location - judicial capital in Indian state} & 154 & 73 & 290 & 69 \\
\hline
{\it\bf owner of a company} & 3103 & 1751 & 4946 & 1053 \\
\hline
{\it\bf location of a parent company} & 0 & 0 & 0 & 1 \\
\hline
{\it\bf location of producer} &11 & 5 & 27 & 8 \\
\hline
{\it\bf place lived by person} & 422 & 260 & 624 & 416 \\
\hline
{\it\bf location - capital of a country } & 150147 & 67047 & 271894 & 59342 \\
\hline
{\it\bf locations - countries within states or provinces} & 0 & 1 & 2 & 0 \\
\hline
{\it\bf person's place of death} & 501 & 286 & 592 & 146 \\
\hline
{\it\bf advisors of company} & 7092 & 2069 & 16879 & 4848 \\
\hline
{\it\bf nationality of person} & 173 & 136 & 728 & 226 \\
\hline
{\it\bf place where the company founded} & 46606 & 23620 &  85529 & 21813 \\
\hline

\end{tabular}
\end{tiny}
\end{center}
\caption{\label{tab:riedeltypes}  Statistics on the relation types per domain.}
\end{table}

\subsection {Entity-Relation Graph, Word Cloud Visualizer and End-to-End Framework Demonstration}

Finally, we evaluated the visualization interface, as perceived by users. To this end, we applied the whole framework — from searching relevant pages for entities, performing relationship extraction, constructing graph-based profiles through to visualizing the end result. In a study carried out at the IUI ’19 conference, we generated entity graphs using the conference participants as the entities of interest and presented them with their own profiles. We then collected their feedback regarding the accuracy and coverage of the facts included in the graph. We further asked for their assessment of the multifaceted graph visualization component of the framework to determine whether it conveyed the relevant details clearly and supported information exploration. In prior work ~\cite{umapamal2020}; ~\cite{amal2020personalized}; ~\cite{amal2020demonstrating}, a high-level description of the study was described with initial results; here we provide more detailed explanations about the visualization we used and the study, detailed statistics and thorough analysis of the study and its collected feedback comments.

\subsubsection{Dataset construction and Demonstration of the whole framework}
To demonstrate and evaluate our framework end-to-end, we constructed a dataset as follows. The names of 600 IUI’19 conference participants and program committee members that appeared on the conference Website were selected. For each of the 600 scholars, we queried Google using their names and picked the top 30 pages. The relevant page finder component was applied on the pages retrieved from Google (4554 relevant Web pages for all 600) to decide which of them are relevant to the entity of interest. Fourth, the entity finder was applied to the relevant pages to extract and label entities that are related to the entity of interest. 33849 related entities were found. The relation finding component was applied to the found related entities, in order to extract and label the relationships between the entity of interest and the related entities. 97447 relationships were found. An entity-relation graph was constructed for each of the 600 entities of interest. In total, for all the 600 entities, 34449 entities and 97447 relationships  were found and used to construct a graph-based profiles for the scholars. Note that relationships can belong to multiple types; hence for Publication, we had 79803 examples, for Employment, we had 60932 examples, and Education, we had 10139 examples.

\subsubsection{Experimental Setup}
\begin{figure*}[t]
\centering
\centering
\begin{tiny}
\includegraphics[width=4cm]{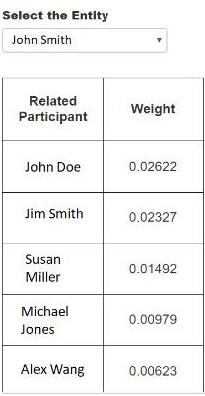} 
\caption{The ranked list presentation.}
\label{fig:rankedlist}
 \end{tiny}
\end{figure*}

Given the generated dataset of the 600 IUI '19 conference participants and program committee members, the constructed profiles were presented using multifaceted graphs (illust rated in Figure ~\ref{fig:prfileviz}) and word cloud visualization (illustrated in Figure ~\ref{fig:wordcloud}) as described in Section 3.3 and made available to the 600 entities of interest. A Web-based system was developed for this purpose and a total of 64 users participated in the study by exploring the system.
We gave users a brief description about the system and how to use it for exploring their generated profile. To acquaint users with the elements of the graph visualization and filtering options, we asked them first to track, in the graph, about three entities of different types, and then to explore their profile. For comparison purposes, we presented users with the same information in the form of a ranked list (as shown in Figure ~\ref{fig:rankedlist}). Following their interaction with the system, users were then asked to fill out a form, which asked for structured feedback regarding the multifaceted information included in the graph, and the accuracy and coverage of the facts included in the profile. Specifically, users were asked to indicate their:
\begin{itemize}
\item preference with respect to the entity-relation graph presentation of the personal profile vs. a ranked list of related entities. 
\item preference with respect to the type of explanation provided regarding the type of relation with each related entity: as a word cloud or as text that surrounded (appearing before and after) the entity mention in the source page. 
\item rating of each filtering control - how helpful/useful each filtering control was during the study.
\end{itemize}

In addition, we asked users for free-form comments about what they liked or disliked in the graph presentation, as well as suggestions for improvements. Finally, we collected implicit feedback, through the system usage log reflecting the number of clicks on the filters and links in the graph. Regarding demographic information, since we ran the user study at a scientific conference during breaks, we did not want to bother the participants with more questions about demographic information. Nevertheless, we know that participants were practitioners, students, or professors.

\subsubsection{Results}

The majority of participants (40 out of 64) preferred the entity-relation graph visualization over the ranked list. Some users (8 out of 64) preferred the ranked lists as they were mostly interested in seeing the top connected entities. Interestingly, a larger group (16) liked both presentations equally as it allowed them more comprehensive exploration. 
As for the explanation of the relation type, 23 users preferred the word cloud while 16 preferred the textual presentation of the surrounding text. The remaining 25 preferred both equally, as it allowed them to first get a highlight of the main concepts and then access the surrounding text upon request.
We also tracked and logged the activities of the users and collected statistics on the clicks on each element of the system during the study. Table ~\ref{tab:clicksstat} presents the click statistics for the filter types mentioned in Section 3.3, in total and on average. As we can see, the filter type with the highest number of clicks is the filter of the number of nodes to be presented in the graph, as well as the relation type filters. We conclude that it was important to users to control the scope of information shown in the graph in terms of number of nodes. Indeed, limiting the number of nodes allows the user to focus on the top connected nodes, and generally allows a high-level overview of the information. Specifying the type of relation of interest allows users to explore the information in a guided fashion.

\begin{table}[t]
\begin{center}
\begin{small}
\begin{tabular} {lll}
\hline
      {\it\bf Name }  & {\it\bf \#clicks Average } & {\it\bf  \# clicks} \\
      
\hline
{\it\bf Relations filter} &   440 & 55 \\

\hline
{\it\bf Nodes filter } & 296 & 49 \\

\hline
{\it\bf \#Nodes filter  } & 612 & 87 \\

\hline
{\it\bf Source of info filter} & 159 & 80 \\

\end{tabular}
\end{small}
\end{center}
\caption{\label{tab:clicksstat}  Click statistics.}

\end{table}

Considering clicks on edges, we saw that a total of 509  clicks on edges were made. We found that the largest number (268) of nodes reached through the clicked edges were of the Person type. Specifically, these persons were found on Web pages other than user’s homepages (187). That is, the multifaceted graph visualization ignited the interest of the users and got them exploring what the system could find about them beyond their homepage information.
Once the users were done exploring and using the system, we asked them to rate to what extent and how helpful each filtering control was, as well as express their opinion regarding the accuracy and coverage of the explored relations. Table ~\ref{tab:userrates} shows the user ratings (1\textendash5 scale). As shown, the ratings are generally high, especially with respect to accuracy and coverage.

\begin{table}[t]
\begin{center}
\begin{small}
\begin{tabular} {ll}
\hline
      {\it\bf Name  }  & {\it\bf Average rate (out of 5)} \\
      
\hline
{\it\bf Relations filter } & 4.13\\
\hline
{\it\bf Nodes filter } &    3.95 \\
\hline
{\it\bf \# Nodes filter } &    3.98 \\
\hline
{\it\bf Information accuracy and coverage } &   4.34 \\
\hline
{\it\bf Temporal aspect } &    3.59 \\
\hline
{\it\bf Temporal aspect } &    3.58 \\
\hline
{\it\bf Publications } &    3.5 \\

\end{tabular}
\end{small}
\end{center}
\caption{\label{tab:userrates}  User ratings.}

\end{table}

The visualization aspects also received high scores. Below, we detail the comments provided by users on the specific aspects that they liked or did not like, along with their suggestions for improvement. 

\begin{itemize}
\item Examples of general users’ comments: 
\begin{itemize}
\item Positive: 

\begin{itemize}
\item Coverage and accuracy:  “Covers broad connections of my career”, “Finding out about new connections”, “The data was relevant to me and seemed to come from multiple sources”, “I liked how it showed my connections and relevant people”, “Graph structure gives a great overview”, “Graph structure very useful for browsing/exploring”, “Compact representation of my career”.
\item Ease of use:  “Easy to grasp an overview/context of a researcher”. “Easy to see an entire relation”. “Allows further investigation if an interesting relation is found”. “I liked the collection in a single graph visualization”. “Ability to drag and move nodes to arrange how I want”. “Very dynamic because of the adaptive scale”, “Main info is on the first page graph. It’s informative”. “The filtering system was helpful in reducing the amount of information that I can follow”. “Liked the adaptive and interactive nature where we can drag, rearrange, and separate out nodes of interest”. “I like the swipe bar for controlling the top entities”.
\end{itemize}
\item Negative: 
\begin{itemize}
    \item Crowdedness: “Gets too busy with many nodes”, “Probably the crowdedness”. “Too many links”, “Edges overlap when graph is full”. 
    \item Technical limitations: “Changing filters repaints the whole graph”. “Could not change the manner in which things clustered”. “The graph wasn't very static between interactions; it often rebuilt and nodes moved from where I had positioned them in an earlier interaction step”. 
\end{itemize}
 
\end{itemize}

\item Comments about the coloring of nodes and edges: 
\begin{itemize}
\item Positive: “Coloring was helpful”. “The colors and shapes helped me easily distinguish different information”. “The encoding with colors was very clear to me”. “The shapes and colors were helpful because I could make immediate sense of the complicated graph”.
\item Negative: “The coloring of edges was less helpful”. “Excessive use of color as visual encoding made it hard for me”. “Color coding for temporal aspects is less useful if all the information is very recent”. “The color of unclassified edges is too bright”. 

\end{itemize}

\item Comments about the word cloud: 
\begin{itemize}
\item Positive: “Word cloud showing some information about the connection”. “Liked the tag/word cloud”. “Word clouds are very interesting”.
\item Negative: Negative: “The word cloud and textual representation are both needed to understand relations”. 
\end{itemize}

\item Comments about node shapes: 
\begin{itemize}
\item Positive: “Preferred the facet of shapes of nodes and width”. “Liked the shapes”. “The colors and shapes are well used and make sense”.
\item Negative: “Rectangles and squares are too similar, hard to understand”. “Didn’t like the node size”. “Width and colors of edges are confusing”. 
\end{itemize}

\item Suggestions for improvements we got from users: 
\begin{itemize}
\item Organization of graphs: “I would use organization of nodes in space as a means of encoding date”. “Organize the graph visualization so the edges are not occluded by the nodes (using another layout algorithm)”. “Use a multiple coordinate view system”.
\item Visualizations of facets: “The recency coloring is better used on edges as it is edge related, not node related”. “Fine-tuning the temporal window with colors would be helpful to researchers who are 25 vs. 70”. “Change node size depending on weights”; “Perhaps providing the nodes in a hierarchical manner could be useful”. 
\item Additional technical capabilities: “Better use of colors as a visual encoding variable”. “It would be nice to use a dynamic social network to explore my profile”. “To be able to search for a person”. “Optional clustering based on types (people vs. objects/organizations)”. “To be able to mark nodes for separate exploration”. “Node interval should be longer”. “This system is interesting in terms of grasping the overview of a person, while I suggest that the system offer a little bit more detailed explanation”. “Click on node instead of edge”. “Filter on years”. “For visualization, allow users to choose which variable the color should represent, as color makes it easier to distinguish nodes in the graph”.

\end{itemize}

\end{itemize}

\subsubsection{Improved Visualization}

\begin{figure*}[t]
\centering
\centering
\begin{small}
\includegraphics[width=13cm]{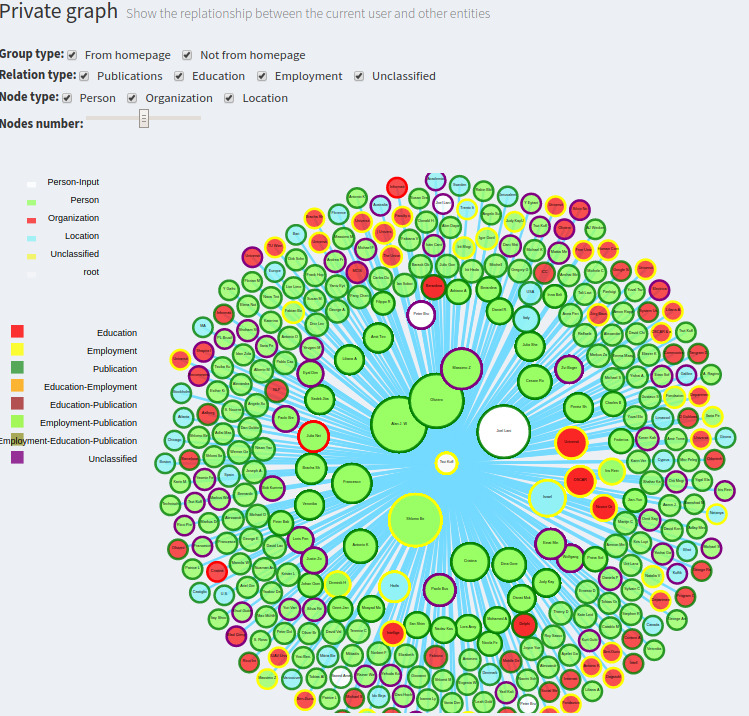} 
\caption{Example of our generated profile visualized using a 2D graph.}
\label{fig:prfileviz_new}
 \end{small}
\end{figure*}

In response to the comments we received from the study conducted at IUI '19, we improved the visualization as shown in Figure ~\ref{fig:prfileviz_new}. The various components of the structured profile are now displayed using a dedicated visualization technique. Specifically,

\begin{itemize}

\item The color inside the node  denotes the type of the nodes with bright green for {\it Person}, red for {\it Organization} and bright blue for {\it Location} and yellow for Unclassified.

\item The color of the node's scope denotes the relation type: dark green color for {\it Publications} relation, red for {\it Education}, yellow for {\it Employment} and dark purple for {\it Unclassified relations}, orange for those classified both as {\it Education} and {\it Employment}, dark brown for those classified both as {\it Education} and {\it Publications}, bright green for those classified both as {\it Employment} and {\it Publications}, and bright brown for those classified as {\it Employment}, {\it Publications} and {\it Education}.

\item The strength of the relation is denoted by both the size of the node, which is linear to the number of the links between them, and the closeness, which is based on 1/(size of node)$^{3}$.
\end{itemize}

\section{Discussion}
\label{sec:discussion}

We have proposed, demonstrated and evaluated a novel framework for person entity profiling, based on fact extraction from relevant Web pages, and the presentation of this structured information in the form of a relational and typed graph. We have focused on extracting and constructing the graphs of person entities and on implementing the framework (with no type limitation on the related entities). We believe that the proposed framework may be generalized to any type of entity of interest. However,this will require further research, in order to explore the benefits and challenges involved in constructing structured graph profiles for others types of entities, such as organizations and locations.

We developed a prototype that searches, analyzes, consolidates and visualizes relevant information that is publicly available on the Web about the entity of interest.

Our first task addressed the problem of identifying a set of relevant Web pages for a specified entity name (person names as our case study), to be used as information sources about that entity. We achieved strong performance (precision of 0.9 and recall of 0.93 (F1 of 0.92))) on this task using a heterogeneous collection of Web pages for persons from different domains. We have showed that high performance may be reached using the top five most informative features (F1=0.87).

We further evaluated the extent to which entities that are related to an entity of interest may be tracked using the various Web pages identified as relevant to this person using our approach, beyond the named entities that appear on their homepage. We evaluated this aspect across domains, and found that we were generally able to identify related entities that were not included on the entity of interest’s homepage. Considering the entities that appear on the entity of interest’s Wikipedia page (and not being included on their homepage) as valid, we found an additional 14\% related entities by exploring relevant Web pages on the Web, beyond the entity of interest’s homepage.

We believe that this work thus shows the feasibility and potential of enhancing entity search by tracking diverse relevant semi-structured information available on the Web. In contrast, existing research typically focuses on structured and well-defined sources such as Wikipedia, which are not available for all persons or all entities.

Admittedly, searching unstructured sources may lead to the need for entity disambiguation, as well as possible concerns regarding the quality of the information and the validity of the underlying source.

Second, we addressed the task of identifying the relations that hold between the entity of interest (person) and the found related entities. Specifically, we relied on the surrounding text of the named entity mentioned in the underlying Web source and automatically categorized it using a set of predefined domain-specific relation types. We performed two studies to evaluate this task. The first study concerned the academic domain, where we defined a small set of relation types (e.g., publications, employment). We used weak supervision in learning to associate entity mentions with the respective types. We developed a dedicated distant supervision scheme for this purpose, automatically assigning entity labels based on the sections in which they appeared in the entity of interest’s CV. Our classification results of unlabeled instances were high, reaching performance of roughly 0.9 in precision and recall across relation types. In another study, we targeted the classification of relations into 32 relation types in several domains. We achieved precision of \textasciitilde0.73 and recall of \textasciitilde0.69 (F1 \textasciitilde0.71) on average, and consider these results to be encouraging, in view of the higher difficulty of this setup. We believe that learning relation types using annotated data on a larger scale will lead to substantial improvement. This may be achieved using distant labeling schemes, such as our CV-mapping labeling approach, while extending it to additional domains.

Next, we addressed the task of constructing a relational and structured graph profile, representing the factual information extracted about the entity of interest, comprising the found related entities and the typed relationships with these related entities. To convey the wealth of information included on a profile clearly and effectively, we implemented various visualization elements, e.g., denoting the related entity types by shapes and the strength of relationship by edge width. In addition, we enable the user to limit the number of objects in the visualized graph, enabling them to focus on the most important related entities according to their edge weights; to limit the object by relation types; and, to limit the viewed information temporally, based on available temporal information.

The various elements of the proposed interface, as well as the graph presentation approach as a whole, were evaluated in several studies at academic conferences, and mainly at IUI’19 which we report in this work. The prototype presented to users received mostly positive feedback. The study participants perceived the accuracy and coverage of the information included on their personal graph as high (assigning an average score of 4.34 out of 5). Importantly, we have shown that the graph presentation was engaging, based on the measured number of clicks. Moreover, according to user feedback, the graph presentation was informative and helpful (averaging a score of 3.54 out of 5).

This evaluation of user experience is in line with our initial assumptions and related research, by which factual structured information that is presented in graphical as opposed to textual form is highly appropriate for entity search. While our user study focused on the scholar domain, we believe that our findings apply to many other domains.

In addition to the limitations noted above in the descriptions of the individual tasks, other limitations that may be removed in future research pertain to the scope of our experiments. As our study involved real participants recruited onsite at academic conferences, this inherently limited the number of participants, as well as the time available per user, which was restricted to conference breaks.

\section{Conclusions and Future Work}
\label{sec:concfuture}

This research presents an integrative novel framework for person entity profiling. Whereas current entity search approaches present the user with lists of results (Web pages) about an entity, we extract relevant factual information about the entity from the Web, and present the user with relational information about the entity of interest in the form of related entities and their interrelationships. We have focused on extracting and constructing the graphs of person entities and on implementing the framework (with no type limitation on the related entities). The proposed framework may, however, be generalized to any type of entity of interest. Furthermore, our framework is not limited to entities included in Wikipedia or other concrete resources. It is applicable to any entity present on the Web.

We have developed a prototype system, which given a query about an entity of interest, searches for relevant Web pages, extracts typed related entities from the Web pages it finds, and consolidates and visualizes (in a graph) relevant public information on the Web about the entity of interest. Importantly, users are provided with intuitive and flexible access to the information contained in the graph, allowing them to receive integrated and rich contextual information about the searched entity, encapsulated in the multifaceted and typed entity-relation graph. Users may readily explore the entity spaces related to the entity in question by following semantic relations between the focus entity and its related entities, at various levels of granularity.

We evaluated the individual components of the framework using datasets that we constructed for this purpose, and demonstrated high performance.

For future work, we would like to address the challenge of entity disambiguation, e.g., tackling fact extraction for persons whose names are common, and may, therefore, map to multiple persons. We also plan to explore the potential and challenges involved in constructing structured graph profiles for various types of entities, such as organizations and locations. Finally, we believe that our framework is useful for multiple tasks - for example, for the automatic generation of a CV document. Future research may explore a variety of applications that can benefit from this framework.

\section{Acknowledgements}
We thank Dr. Zef Segal and Mr. Mustafa Adam for their contribution to this research



\bibliographystyle{model5-names}
\bibliography{main}




\end{document}